\documentclass[9pt,twocolumn,twoside]{pnas-new}
\templatetype{pnasresearcharticle} % Choose template 
\usepackage{graphicx}% Include figure files
\usepackage{dcolumn}% Align table columns on decimal point
\usepackage{bm}% bold math
\usepackage{todonotes}
\usepackage{amsmath}	% required for `\align' (yatex added)
\let\a=\alpha \let\b=\beta  \let\d=\delta
   
\let\l=\lambda \let\m=\mu  \let\x=\xi 
\let\s=\sigma   
   
\let\D=\Delta   
   
\let\ee=\varepsilon \let\r=\rho \let\th=\theta \let\io=\infty

\def\NN{{\cal N}}

\def\to{\rightarrow}

\newcommand{\beq}{\begin{equation}} \newcommand{\eeq}{\end{equation}}

\newcommand{\pdiff}[2]{\frac{\partial #1}{\partial #2}}

\newcommand{\new}{\nonumber\\}
\newcommand{\abs}[1]{\left|#1\right|}

\newcommand{\hr}{\hat{\bm{r}}}
\newcommand{\hu}{\hat{\bm{u}}}

\newcommand{\bx}{\bm{x}}
\newcommand{\bxi}{\bm{\xi}}
\newcommand{\ave}[1]{\left\langle #1 \right\rangle}

\title{Universality of jamming of non-spherical particles}

\author[a]{Carolina Brito} \author[b,1]{Harukuni Ikeda}
\author[c]{Pierfrancesco Urbani} \author[d]{Matthieu Wyart}
\author[b]{Francesco Zamponi}

\affil[a]{Instituto de F\'isica, UFRGS, 91501-970, Porto Alegre, Brazil}
\affil[b]{Laboratoire de physique th\'eorique, D\'epartement de physique de l'ENS, \'Ecole normale sup\'erieure, PSL University, Sorbonne Universit\'e, CNRS, 75005 Paris, France}
\affil[c]{Institut de physique th\'eorique, Universit\'e Paris Saclay, CNRS, CEA, F-91191 Gif-sur-Yvette, France}
\affil[d]{Institute of Physics, EPFL, CH-1015 Lausanne, Switzerland}

\leadauthor{Brito} 

\significancestatement{ 
The jamming transition is a key property of
granular materials, including sand and dense suspensions.
In the generic situation of
non-spherical particles, its scaling
properties are not completely understood.
Previous empirical and theoretical work in ellipsoids and
spherocylinders indicates that both structural and vibrational properties
are fundamentally affected by shape.  Here we explain these observations
using a combination of marginal stability arguments and the replica
method. We unravel a new universality class for particles with internal
degrees of freedom, and derive how the structure of packings and their
vibrations scale as the particles evolve toward spheres.  
}

\authorcontributions{B.C., H.I., P.U., M.W. and F.Z. designed research,
performed research, and wrote the paper.}
\authordeclaration{The authors declare no conflict of interest.}
\correspondingauthor{\textsuperscript{1}To whom correspondence should be addressed.
E-mail: harukuni.ikeda@lpt.ens.fr}


\begin{abstract} %MAX 250 WORDS
 Amorphous packings of non-spherical particles such as ellipsoids and
spherocylinders are known to be hypostatic: the number of mechanical
contacts between particles is smaller than the number of degrees of
freedom, thus violating Maxwell's mechanical stability criterion. In
this work, we propose a general theory of hypostatic amorphous packings
and the associated jamming transition.  First, we show that many systems
fall into a same universality class. As an example, we explicitly map
ellipsoids into a system of ``breathing'' particles.  We show by using a
marginal stability argument that in both cases jammed packings are
hypostatic, and that the critical exponents related to the contact
number and the vibrational density of states are the same.  Furthermore,
we introduce a generalized perceptron model which can be solved
analytically by the replica method. The analytical solution predicts
critical exponents in the same hypostatic jamming universality class.
Our analysis further reveals that the force and gap distributions of
hypostatic jamming do not show power-law behavior, in marked contrast to
the isostatic jamming of spherical particles. Finally, we confirm our
theoretical predictions by numerical simulations.
\end{abstract}

\dates{This manuscript was compiled on \today}
\doi{\url{www.pnas.org/cgi/doi/10.1073/pnas.XXXXXXXXXX}}

\begin{document}
% Optional adjustment to line up main text (after abstract) of first page with line numbers, when using both lineno and twocolumn options.
% You should only change this length when you've finalised the article contents.
\verticaladjustment{-2pt}

\maketitle
\thispagestyle{firststyle}
\ifthenelse{\boolean{shortarticle}}{\ifthenelse{\boolean{singlecolumn}}{\abscontentformatted}{\abscontent}}{}

% If your first paragraph (i.e. with the \dropcap) contains a list environment (quote, quotation, theorem, definition, enumerate, itemize...), the line after the list may have some extra indentation. If this is the case, add \parshape=0 to the end of the list environment.

\dropcap{U}pon compression, an athermal system consisting of purely
repulsive particles suddenly acquires finite rigidity at a certain
jamming transition density $\varphi_J$ at which constituent particles
start to touch each other producing a finite mechanical pressure~\cite{Liu10,van2009,torquato2010}.
The jamming transition is observed in a wide variety of
physical, engineering and biological systems such as metallic
balls~\cite{bernal1960}, foams~\cite{durian1995,PhysRevE.68.011306},
colloids~\cite{zhang2009}, polymers~\cite{karayiannis2009},
candies~\cite{donev2004}, dices~\cite{jaoshvili2010} and
tissues~\cite{bi2015}. In the past decade, a lot of progress has been
made in understanding the jamming transition of spherical and
frictionless particles with repulsive interactions. Key findings involve
(i) the power law behaviors of the elastic modulus and contact number as
a function of the proximity to
$\varphi_J$~\cite{ohern2002,PhysRevE.68.011306,PhysRevLett.97.258001},
(ii) the emergence of excess soft modes in the vibrational density of
states $D(\omega)$~\cite{PhysRevE.68.011306}, and (iii)~the power law
divergence of the gap distribution function $g(h)$ and power law tail of
the force distribution function $P(f)$ at
$\varphi_J$~\cite{PhysRevE.68.011306,PhysRevE.71.011105,charbonneau2014fractal,PhysRevLett.114.125504}.
Those phenomena can be understood in terms of a marginal stability
principle~\cite{PhysRevE.72.051306,wyart2005rigidity}:
the system lies close to a mechanical instability. More precisely, at
$\varphi_J$, the contact number per particle is
$z_J=2d$~\cite{bernal1960,PhysRevE.68.011306}, which barely satisfies
the Maxwell's mechanical stability
condition~\cite{maxwell1864}. Accepting marginal stability as a basic
principle, one can successfully predict the critical exponents of soft
spheres~\cite{PhysRevE.72.051306,wyart2005rigidity}
and derive a scaling relation between critical exponents of hard
spheres~\cite{brito2006,PhysRevLett.109.125502,lerner2013low,degiuli2014}.
The importance of marginal stability is also highlighted by exact
calculations for hard spheres in the large dimension
limit~\cite{charbonneau2014fractal} and in a perceptron model of the
jamming transition~\cite{franz2015,franz2016,franz2017}.  These first
principle calculations prove that a full replica symmetric breaking
(RSB) phase transition occurs ahead of the jamming transition. In the
full RSB phase, the eigenvalue distribution function is gapless, and
thus, the system is indeed marginally stable~\cite{franz2015}. This
approach provides exact results for the critical exponents, which agree
well with the numerical results~\cite{charbonneau2014fractal}, once
localized excitation modes are carefully
separated~\cite{lerner2013low,PhysRevLett.114.125504}.

However, a system of spherical particles is an idealized model and, in
reality, constituent particles are, in general, non-spherical. In this
case, one should specify the direction of each particle in addition to
the particle position. The effects of those extra degrees of freedom have been
investigated in detail in the case of
ellipsoids~\cite{donev2004,man2005,delaney2005,donev2007,mailman2009,zeravcic2009,schreck2012,van2009,torquato2010}.
Notably, the contact number at the jamming point continuously increases
from the isostatic value of spheres, as $z_J-2d\propto \Delta^{1/2}$,
where $\Delta$ denotes the deviation from the perfectly spherical shape.
The system is thus {\it hypostatic}: the contact number is
lower than what expected by the naive Maxwell's stability condition, which would predict
$z_J = 2(d+d_{\rm ex})$ where $d_{\rm ex}$ is the number of rotational
degrees of freedom per particle~\cite{donev2004,delaney2005,donev2007}.
As a consequence of hypostaticity, $D(\omega)$ has anomalous zero modes at
$\varphi_J$, which are referred to as ``quartic modes'' because they are
stabilized by quartic terms of the potential
energy~\cite{donev2007,mailman2009,zeravcic2009,schreck2012}. 
Hypostatic packings are also obtained for
spherocylinders~\cite{williams2003,blouwolff2006,wouterse2007,wouterse2009,teitel2018},
superballs~\cite{jiao2010},
superellipsoids~\cite{delaney2010}, other convex shaped
particles~\cite{werf2018} and even deformable
polygons~\cite{boromand2018}.  Compared to spherical particles, the
theoretical understanding of the jamming transition of non-spherical
particles is still in its infancy~\cite{donev2007,baule2013}. In
particular, the physical mechanism that induces a scaling behavior such as
$z_J-2d\propto \Delta^{1/2}$ is unclear.

In this work, we propose a theoretical framework to describe the
universality class of hypostatic jamming.  As a first example of
universality, we will map ellipsoids into a model of ``breathing''
spherical particles (BP), recently introduced in~\cite{brito2018}.
Based on the mapping, we show that the two models indeed have the same
critical exponents by using a marginal stability argument. Next, we
propose a generalisation of the random perceptron model that mimics the
BP and can be solved analytically using the replica method. We confirm
that this model is in the same universality class of ellipsoids, BP, and
other non-spherical particles that display hypostatic jamming.  This
analysis further predicts the scaling behavior of $g(h)$ and $P(f)$ near
the jamming point. Interestingly, we find that these functions do not
show a power-law behavior even at the jamming point, in marked contrast
to the jamming of spherical particles.  Also the simplicity of the model
allows us to derive an analytical expression of the density of states
$D(\omega)$, which exhibits the very same scaling behavior of that of
ellipsoids and BP.  Finally, we confirm our predictions by numerical
simulations of the BP model.

\paragraph*{Breathing particles model --}
The BP model~\cite{brito2018} was originally introduced to understand the
physics of the Swap Monte Carlo
algorithm~\cite{PhysRevX.7.021039},
but here we will focus on its relation with the jamming
of ellipsoids. The model consists of $N$
spherical particles with positions $\bx_i$ in $d$-dimensions 
and radius $R_i \geq 0$, 
interacting via the potential energy:
\begin{align}
V_N(\{\bx\},\{R\}) =U_N(\{\bx\},\{R\}) + \mu_N(\{R\}) \ ,
\label{144601_17May18} 
\end{align}
where, defining $\th(x)$ as the Heaviside theta function,
\begin{align}
 U_N &= \sum_{i<j}k\frac{h_{ij}^2}{2}\theta(-h_{ij}) \ , \quad h_{ij} = \abs{\bx_i-\bx_j}-R_i-R_j  \ ,
\end{align}
is the standard harmonic repulsive interaction potential of spherical particles such as
bubbles and colloids~\cite{durian1995}, 
and the distribution of $R_i$,
which can fluctuate around a reference value $R_i^0$,  
 is controlled by the chemical potential term:
\begin{align}
 \mu_N &= \frac{k_R}{2}\sum_i (R_i-R_i^0)^2 \left(\frac{R_i^{0}}{R_i}\right)^2 \ .
 \label{155329_18May18}
\end{align}
Here, $k_R$ is determined by imposing that the dimensionless standard
deviation $\D \propto \sqrt{\sum_{i}(R_i-R_i^{0})^2/(N R_0^2)}$ is
constant, with $R_0 = N^{-1}\sum_i R_i^0$.  Note that $\D=0$
(corresponding to $k_R=\io$) gives back the usual spherical
particles~\cite{durian1995}, and that the full distribution of radii,
$P(R)$, can generically change even if $\D$ is kept fixed.  Upon
approaching jamming, where the adimensional pressure $p$ (in units of $k
R_0^{2-d}$) vanishes, it is found that $k_R = p/\Delta$ and $P(R)$
remains constant~\cite{brito2018}.

Because the BP model has $Nd$ translational degrees of freedom
and $N$ radial degrees of freedom, the naive Maxwell
stability condition requires $z\geq 2(d+1)$ in the thermodynamic limit~\cite{maxwell1864,alexander1998}. 
However, a marginal stability
argument and numerical simulations prove that the contact number at the
jamming point $z_J$ increases continuously as $z_J-2d \propto
\Delta^{1/2}$~\cite{brito2018} and the system is hypostatic for
sufficiently small $\Delta$, {\it i.e.}, the number of constraints is
smaller than that required by the Maxwell's stability condition. This is
very similar to ellipsoids and motivates us to conjecture
that the two models could belong to the same universality
class. In the following, we show that this expectation is
indeed true: hypostatic packings of the BP and ellipsoids are
stabilized by a common mechanism and have the same critical exponents.

\paragraph*{Mapping from ellipsoids to BP --}
We now construct a mapping from a system of ellipsoids to the
spherical BP model introduced above. 
Ellipsoids are described by their position $\bx_i$ and by
unit vectors $\hu_i$ along their principal axis, and 
for concreteness, we model them by the Gay-Berne
potential~\cite{gay1981,zeravcic2009}:
\begin{align}
 V_N(\{\bx\},\{\hu\}) = \sum_{i<j}v(h_{ij}) \ , \ \ \   v(h)=  k \frac{h^2}{2}\theta(-h) \ ,
 \label{134718_15May18}
\end{align}
where the gap function is defined as
\begin{align}
 h_{ij} &= \frac{\abs{\bx_i-\bx_j}-\sigma_{ij}}{\sigma_0} \ ,
 \new
 \frac{\sigma_{ij}}{\sigma_0} &=
 \frac{1}{\sqrt{1-\frac{\chi}{2}
 \left(
 \frac{\left(\hr_{ij}\cdot\hu_i + \hr_{ij}\cdot\hu_j\right)^2}{1+\chi\hu_i\cdot\hu_j }
 + \frac{\left(\hr_{ij}\cdot\hu_i - \hr_{ij}\cdot\hu_j\right)^2}{1-\chi\hu_i\cdot\hu_j }
 \right)}} \ .
\end{align}
Here, $\hr_{ij} = (\bx_i - \bx_j)/\abs{\bx_i-\bx_j}$ is the unit vector
connecting the $i$-th and $j$-th particles, $\ee\s_0$ is the length of
the principal axis, and $\chi=(\varepsilon^2-1)/(\varepsilon^2+1)$,
where $\varepsilon$ denotes the aspect ratio. Because we are interested
in the nearly spherical case, we expand the pair potential in small
$\Delta= \varepsilon-1$ as
 \begin{align}
  v(h_{ij}) &= v(h^{(0)}_{ij})
  -\frac{\Delta}{2}v'(h^{(0)}_{ij})\left[\left(\hr_{ij}\cdot\hu_i\right)^2
  + \left(\hr_{ij}\cdot\hu_j\right)^2\right]\new
  &+ \Delta^2 w_{ij} \ ,
 \end{align}
where $h_{ij}^{(0)} =r_{ij}/\sigma_0-1$ and $\Delta^2 w_{ij}$ denotes the
$O(\Delta^2)$ term that we do not need to write explicitly. 
Substituting this in Eq.~(\ref{134718_15May18}) and keeping terms up to 
$\Delta^2$, we obtain
 $V_N \approx U_N + \mu_N$, 
where
\begin{align}
 U_N &= \sum_{i<j}\left[v(h_{ij}^{(0)})+\Delta^2 w_{ij}\right] \ , \new
 \mu_N &= \frac{1}{2}\sum_{i}(\Delta\hu_i)\cdot k_{i}\cdot(\Delta\hu_i) \ .
 \label{155357_18May18}
\end{align}
The stiffness matrix is $k_{i}^{ab} = -\Delta^{-1} \sum_{j(\neq i)}v'(h_{ij}^{(0)})\hr_{ij}^a\hr_{ij}^b$, where $a,b=1,\cdots, d$. Note
that near the jamming point, $k_i$ behaves as $k_{i}\sim
v'(h)/\D \sim p/\D$, which is the same scaling of the
stiffness $k_R$ of the BP model, Eq.~(\ref{155329_18May18}).  
Hence, if we identify $\Delta\hu_i$ with $R_i$,
in the vicinity of jamming the potential for 
ellipsoids can be analyzed essentially in the same 
way as the BP model~\cite{brito2018}, as we discuss next.

\begin{figure}[t]
\begin{center}
\includegraphics[width=.8\columnwidth]{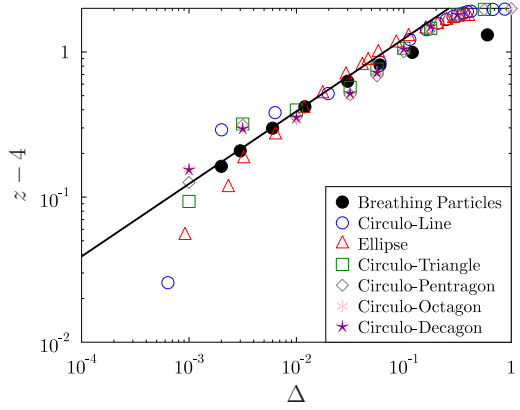} \caption{ {\bf
Universal scaling of the contact number --} Markers denote the numerical
result, while the full line denotes the theoretical prediction $\delta
z\sim \Delta^{1/2}$. Data for non-spherical particles are reproduced
from Ref.~\cite{werf2018}, and from the sphericity $A$, we defined
$\Delta= c(A-1)^{1/2}$, which recovers the correct scaling relation
between the sphericity and aspect ratio of ellipses for small
$\Delta$. We set $c=1/6$ to collapse all data. Data for the BP
correspond to a pressure $p=10^{-6}$. }  \label{fig:dz}
\end{center}
\end{figure}

 \paragraph*{Marginal stability --}
 The distinctive feature of both BP and
ellipsoids is that the total potential, and thus the Hessian matrix,
can be split in two parts: one having finite stiffness, and the second having
vanishing stiffness $p/\D$ by dimensional arguments. The zero modes of the first term
are stabilized by the second, as recognized in Refs.~\cite{donev2007,schreck2012}.
 We now provide additional insight on this structure by generalizing a
 marginal stability argument discussed for the BP in Ref.~\cite{brito2018}. 
 At jamming, $p=0$ and $V_N=U_N$ because $\mu_N \propto p$.
The $\NN_3 \equiv Nz/2$ constraints coming from $U_N$, one per
mechanical contact, stabilize the same number of vibrational modes.
Because the system is hypostatic, there remain $\NN_0 \equiv N(d+d_{\rm
ex})-Nz/2=N(d_{\rm ex}-\delta z/2)$ zero-frequency modes, where $\delta
z=z-2d$ and $d_{\rm ex}$ is the number of extra degree of freedom per
particle, {\it i.e.}, $d_{\rm ex}=1$ for the BP and $d_{\rm ex}=d-1$ for
ellipsoids. Above jamming, where $p>0$, the $\NN_0$ zero modes are
stabilized by the ``soft'' constraint coming from $\mu_N$ whose
characteristic stiffness is $k_R\sim k_i\sim k (p/\D) \ll k$, where $k$
is the stiffness associated to $U_N$.  Hence, the energy scale of these
modes remains well separated from that of the $\NN_3$ other modes, and
we can restrict to the $\NN_0$-dimensional subspace of the soft modes.
In this space, we have $\NN_0 = N(d_{\rm ex}-\delta z/2)$ degrees of
freedom, and $\mu_N$ provides $Nd_{\rm ex}$ constraints, hence the
number of degrees of freedom is $N\delta z/2$ less than the number of
constraints. When $\delta z \ll 1$, a variational
argument was developed in~\cite{PhysRevE.72.051306,yan2016} to describe
the low-frequency spectrum.  It shows that the soft modes are shifted
above a characteristic frequency $\omega_*^2 \sim k_i \delta z^2\sim k_R
\delta z^2 \sim \Delta^{-1} p \, \delta z^2$, which is reduced by $\sim
-p$ by the so-called pre-stress terms, resulting in $\omega_*(p)^2 = c_1
\Delta^{-1}p\delta z^2 -c_2 p$, where $c_1$ and $c_2$ are unknown
constants. Assuming that the system is marginally stable,
$\omega_*(p)=0$, results in~\cite{brito2018}
\begin{align}
 \delta z \sim \Delta^{1/2}.\label{160503_17May18}
\end{align}
This explains the universal square root singularity of the contact
number $z_J$ observed in ellipsoids, BP and several other
models~\cite{donev2004,donev2007,brito2018}, as illustrated in
Fig.~\ref{fig:dz}.
Eq.~(\ref{160503_17May18}) holds when $p\ll \Delta$, because in the
argument we assumed to be close to jamming ($p\sim 0$) at fixed $\D$.
On the contrary, when $\Delta\ll p$, the contact number should have the
same scaling of spherical particles:
\begin{align}
 \delta z \sim p^{1/2} \ .\label{160506_17May18}
\end{align}
Eqs.~(\ref{160503_17May18}) and (\ref{160506_17May18}) imply that $p$
and $\Delta$ have the same scaling dimension and the following scaling holds:
\begin{align}
 \delta z = \Delta^{\gamma}f\left(p/\D\right).\label{161640_17May18}
\end{align}
In the $\Delta\to 0$ limit, Eq.~(\ref{161640_17May18}) reduces to
Eq.~(\ref{160506_17May18}), which requires $\gamma=1/2$ and $f(x)\to
x^{1/2}$ for $x\gg 1$.  In the $p\to 0$ limit, we should recover
Eq.~(\ref{160503_17May18}), which requires $f(x)\to const$ for $x\ll 1$.
For the BP, Eq.~(\ref{161640_17May18}) is confirmed by numerical
simulations~\cite{brito2018}. Assuming that $f(x)$ is a regular function
around $x\sim 0$, one can expand it as $f(x)= c_0 + c_1 x + \cdots$ and
obtains
\begin{align}
z-z_J\sim \Delta^{-1/2}p,\label{155702_18May18}
\end{align}
where $z_J = 2d + c_0\Delta^{1/2}$.
This is compatible with previous numerical
results of ellipsoids, where $z-z_J\sim \Delta^{-0.35\pm
0.1}p$~\cite{schreck2010}.
We can also study the response to shear deformation, which
mainly excites the
zero-modes~\cite{mailman2009}. 
Applying the argument in Ref.~\cite{wyart2005rigidity} to
the zero-modes and using Eq.~(\ref{160503_17May18}),
the shear modulus $G$ behaves as $G\sim \delta z k_R\sim \delta z k_i
\sim p/\sqrt{\Delta}$, in perfectly agreement with the numerical
result~\cite{mailman2009}.

 \paragraph*{Vibrational spectrum --} The marginal stability argument suggests
 that $\NN_0$ soft vibrational modes can be found in the frequency range
 $\omega^* \lesssim \omega \lesssim \sqrt{k_R}$, with $\omega^*\sim 0$
 due to marginal stability and $k_R \sim p/\D$, while the remaining
 $\NN_3$ modes have finite frequency at jamming. We now refine the
 argument to discuss in more details the vibrational density of states
 $D(\omega)$. It is convenient to define the $\NN \times \NN$ Hessian
 matrix of the BP model, with $\NN = N(d+d_{\rm ex})$, as the second
 derivative of the interaction potential $V_N$ w.r.t. $\bx_i$ and
 $R_i/\Delta$, in such a way that it has a similar scaling of the one of
 ellipsoids, where $R_i/\Delta$ is mapped onto the angular degrees of
 freedom $\hu$.

Then, $D(\omega)$ near jamming can be separated into the following three
regions. (i)~The lowest band corresponds to the $\NN_0=N(d_{\rm
ex}-\delta z/2)$ zero modes stabilized by $\mu_N$. Their typical
frequency is $\omega_0^2 \sim \partial^2\mu_N/\partial
(\Delta^{-1}R_i)^2 \sim k_R \D^2 \sim \Delta p$.  The remaining $\NN_3 =
\NN - \NN_0 = Nz/2$ modes can be split into two bands: (ii)~an
intermediate band corresponding to the extra (rotational or radial)
degrees of freedom $\NN_1 = N \d z/2$, with typical frequency
$\omega_1^2 \sim \partial^2 V_N/\partial (\Delta^{-1}R_i)^2 \sim
\Delta^2 $, and (iii)~the highest band corresponding to the $\NN_2 = Nd$
translational degree of freedom.  For $\Delta\ll 1$, the additional
degrees of freedom do not strongly affect these modes, and one can apply
the standard variational argument of spherical
particles~\cite{PhysRevE.72.051306,yan2016}, which predicts that their
typical frequency is $\omega_2^2 \sim \delta z^2 \sim \Delta$. 
The resulting $D(\omega)$ differs significantly from that of isostatic packings of
spherical particles, which displays a single translational band.

\begin{figure*}[t]
\begin{center}
\includegraphics[width=.8\textwidth]{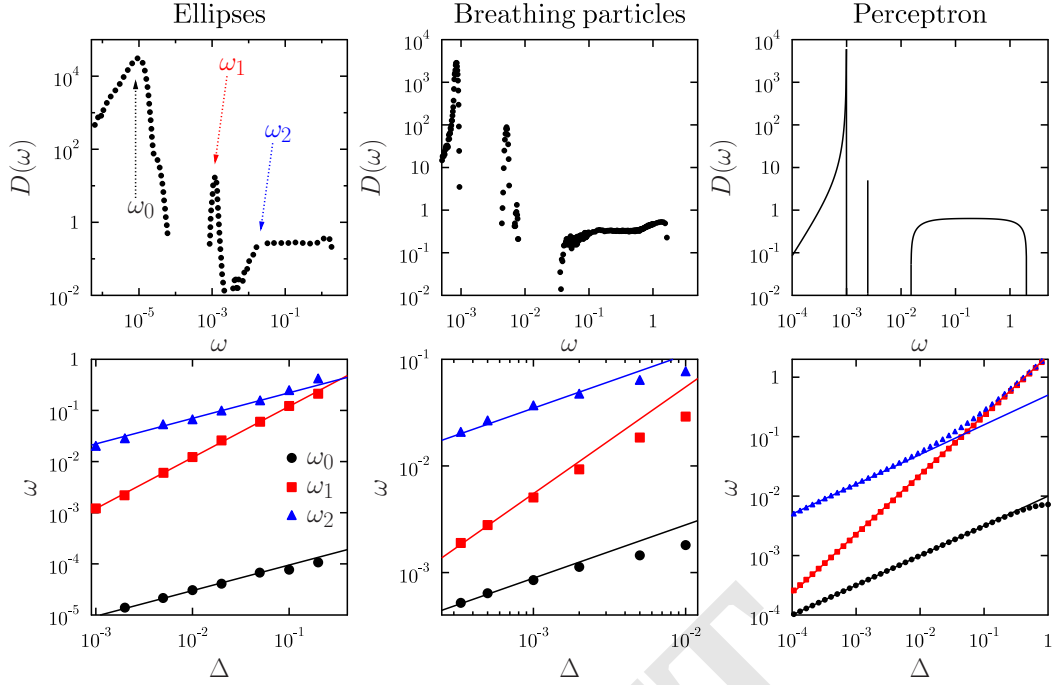} \caption{ {\bf
Universality of the density of states --} (Top) Density of states for
ellipses, breathing particles, and the perceptron.  (Bottom) Evolution
with $\D$ of the characteristic frequencies at $p=10^{-4}$. Full lines
denote the theoretical predictions, $\omega_0\propto \Delta^{1/2}$,
$\omega_1\propto \Delta$ and $\omega_2\propto \Delta^{1/2}$,
respectively.  Data of ellipses are reproduced from
Ref.~\cite{schreck2012}.}  \label{fig:DOS}
\end{center}
\end{figure*}

\begin{figure}[t]
\begin{center}
\includegraphics[width=\columnwidth]{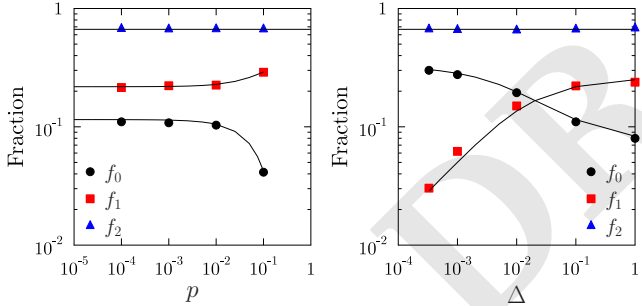} \caption{ 
{\bf Weights of the density of states --} Fraction of modes $f_i = \NN_i/\NN$ in the three bands of $D(\omega)$
given in Fig.~\ref{fig:DOS}, plotted as functions of $p$ at fixed $\D=10^{-1}$ (left) and $\D$ at fixed $p=10^{-4}$ (right) for breathing particles (with $d=2$ and $d_{\rm ex}=1$).
The theoretical predictions $f_0 = (1-\d z/2)/3$, $f_1 = \d z/6$ and $f_2 = 2/3$ are plotted as full lines, inferred from the measured $\d z$.
 }
\label{fig:N}
\end{center}
\end{figure}

Numerical results for $D(\omega)$ of ellipsoids from~\cite{mailman2009},
of the BP from~\cite{brito2018}, and analytical results for the
perceptron model to be introduced below, are reported in
Fig.~\ref{fig:DOS}.  Details about the simulations of the BP are
explained in~\cite{brito2018}; here we show data for $N=484$ particles,
averaged over at least 1000 samples for each state point.  As predicted
by our theory, $D(\omega)$ consists of three separated bands with
characteristic peak frequencies $\omega_{0,1,2}$. Their scaling with
$\D$, also reported in Fig.~\ref{fig:DOS} at fixed $p$, follows the
theoretical predictions $\omega_0\propto \Delta^{1/2}$, $\omega_1\propto
\Delta$ and $\omega_2\propto \Delta^{1/2}$. We also find that
$\omega_0\propto p^{1/2}$ for small $p$, while $\omega_{1,2}$ do not
change much with $p$, which is again consistent with the
theory. Finally, in Fig.~\ref{fig:N} we report the fraction $f_i =
\NN_i/\NN$ of modes in each band for the BP, which also follow the
theoretical prediction as a function of $\D$ and $p$.

\paragraph*{Mean field model --}
The universality class of isostatic jamming is well understood: it can
be described analytically by particles in
$d\to\io$~\cite{charbonneau2014fractal} or, equivalently, by the
perceptron model~\cite{franz2015,franz2016,franz2017}: both models
reproduce the critical exponents of isostatic jamming in all dimensions
$d$, leading to the conjecture that its lower critical dimension is
$d=2$~\cite{goodrich2012finite}.

We now introduce a new mean field model which describes the universality
class of hypostatic jamming in the BP, ellipsoids and many other models
of non-spherical particles.  The model, which is a generalization of the
perceptron, can be solved analytically and, as we shall show, the
solution reproduces all the critical exponents of hypostatic jamming. It
consists of one tracer particle with coordinate $\bx$ on the surface of
the $N$ dimensional hypersphere of radius $\sqrt{N}$, and $M$ obstacles
of coordinates $\bxi_\m$ and ``size'' $\s + R_\m$.  The interaction
potential between the tracer particle and the obstacles is
\begin{align}
 V_N &= U_N + \mu_N \ , \
 U_N = \sum_{\mu=1}^M v(h_\mu) \ , \
 \mu_N = \frac{k_R}{2}\sum_{\mu=1}^MR_{\mu}^2 \ ,
\label{133037_18May18}
\end{align}
where $v(h)=h^2\theta(-h)/2$ and the gap variable $h_\mu$ is defined as
\begin{align}
 h_{\mu} &= \frac{\bx\cdot\bxi_\mu}{\sqrt{N}}-\sigma-R_\mu \ .
\end{align}
The $\bxi_\mu$ are frozen variables, and each of their components
follows independently a normal distribution of zero mean and unit
variance.  The dynamical variables are $\bx$ and the $R_\m$, whose
variance is controlled by the chemical potential $\mu_N$. We fix the
value of $k_R$ so that $\sum_{\mu=1}^M R_\mu^2 = M\Delta^2$. In the
$\Delta\to 0$ limit, the system reduces to the standard perceptron model
investigated in Ref.~\cite{franz2017}, while for $\D>0$ the $R_\mu$ play
the same role of the particle sizes in the BP model.

\begin{figure}[t]
\begin{center}
\includegraphics[width=8cm]{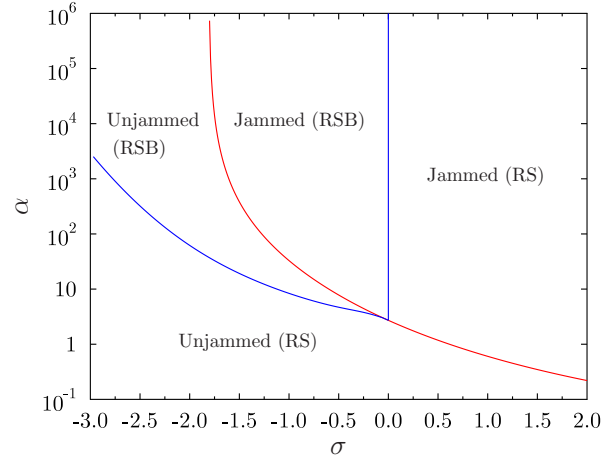} \caption{ The phase diagram of
the perceptron model for $\Delta=0.1$. The red line denotes the jamming
point.  The blue lines denote the RSB instability. The jamming line in the
nonconvex region ($\sigma<0$) is surrounded by the RSB lines.  }
\label{150903_18May18}
\end{center}
\end{figure}

Because the model can be solved by the same procedure of the standard
perceptron model, here we just give a brief sketch of our calculation,
which will be given in a longer publication. The free energy of the
model at temperature $T=1/\b$ can be calculated by the replica method,
$-\beta f = \lim_{n\to 0}\frac{1}{nN}\log \overline{Z^n}$, where $Z=\int
d^N \bx d^MR e^{-\beta V_N}$ and the overline denotes the averaging over
the quenched randomness $\bxi_\mu$. Here we are interested in the
athermal limit $T\to 0$.  Using the saddle point method, the free energy
can be expressed as a function of the overlap
$q_{ab}=\ave{\bx^a\cdot\bx^b}/N$, where $\bx^a$ and $\bx^b$ denote the
positions of the tracer particles of the $a$-th and $b$-th replicas,
respectively. In the $n\to 0$ limit, $q_{ab}$ is parametrized by a
continuous variable $x\in [0,1]$, $q_{ab}\to q(x)$. The function $q(x)$
plays the role of the order parameter and characterizes the hierarchical
structure of the metastable states~\cite{mezard1987}.  We first
calculate the phase diagram assuming a constant $q(x)=q$, which is the
so-called replica symmetric (RS) ansatz that describes an energy
landscape with a single minimum.  The result for $\Delta=0.1$ is shown
in Fig.~\ref{150903_18May18}.  The control parameters are the obstacle
density $\alpha= M/N$ and size $\sigma$.  If $\alpha$ is small, the
tracer particle can easily find islands of configurations $\bx$ that
satisfy all the constraints $h_\mu>0$: the total potential energy $U_N$
and the pressure vanish and the system is unjammed.  The overlap $q<1$
measures the typical distance between two zero-energy configurations.
Upon increasing $\alpha$, $q$ increases and eventually reaches $q=1$ at
$\alpha_J$, which is the jamming transition point
(Fig.~\ref{150903_18May18}).  Naturally, due to the additional degrees
of freedom when $\D>0$, we have $\alpha_J(\Delta)>\alpha_J(0)$ for equal
$\s$.  For $\sigma>0$, the RS ansatz is stable for all values of
$\alpha$ and it describes the jamming transition. For $\sigma<0$
instead, the jamming line is surrounded by a replica symmetry broken
(RSB) region where the RS ansatz is unstable.  The jamming transition
should thus be described by the RSB ansatz where $q(x)$ is not constant,
corresponding to a rough energy landscape.  The qualitative behavior of
the phase diagram is independent of $\Delta$, in particular the jamming
line for $\sigma<0$ is always surrounded by a RSB region.

An important observable to characterize jamming is the gap distribution
$\rho(h)\equiv \frac{1}{N}\sum_{\mu=1}^M \ave{\delta(h_\mu-h)}$ that
also gives the contact number $z=\int_{-\infty}^0 dh \rho(h)$.  At
jamming, $z$ counts the gaps $h_\mu$ that are exactly equal to zero.
For comparison with numerical results, we introduce the positive gap
distribution $g(h) \equiv \th(h) \rho(h)/\int_{0}^\infty dh \rho(h)$,
and the force distribution $P(f) \equiv
\theta(-h)\rho(h)\pdiff{h}{f}/\int_{-\infty}^0\rho(h)\pdiff{h}{f}df$,
where $f=-h/p$ (corresponding to negative gaps), both normalized to 1.
For the standard perceptron model with $\Delta=0$ and $\s<0$, jamming is
isostatic with $z=1$~\cite{franz2017}, and both $g(h)$ and $P(f)$
exhibit a power law behavior~\cite{franz2015,franz2016,franz2017}.  In
the jammed phase and $\alpha\gtrsim \alpha_J$, the system is described
by a ``regular'' full RSB solution where $1-q(x) \sim y_\chi^2 x^{-2}$
for $q(x)\sim 1$, and $g(h)$ and $P(f)$ are regular and finite
functions.  The prefactor $y_\chi$ is predominantly controlled by the
contact number $z$, and diverges at isostaticity when
$z=1$~\cite{franz2017} and the regular solution breaks down.  At $\a_J$,
the model is described by the ``jamming'' solution where $1-q(x)\sim
x^{-\kappa}$, $g(h)\sim h^{-\gamma}$ and $P(f)\sim f^{\theta}$, with
critical exponents $\kappa\simeq 1.42$, $\gamma=(2-\kappa)/\kappa$ and
$\theta=(3\kappa-4)/(2-\kappa)$~\cite{charbonneau2014fractal,franz2015,franz2016,franz2017}.
Near $\alpha_J$, the regular solution should connect to the jamming
solution.  This matching argument leads to $z-1\sim p^{1/2}$, which is
the same scaling behavior of spherical
particles~\cite{PhysRevE.68.011306}.

The situation is completely different if $\Delta>0$. One can show that the
contact number at jamming is $z_J\geq 1$, meaning that the
regular solution persists even at $\alpha_J$.  Consequently, $g(h)$ and
$P(f)$ are finite and regular functions at jamming, and the square-root behavior of 
the contact number is replaced by $z-z_J = c_{\Delta} p$.  
At $\alpha_J$, the regular solution should connect to the jamming
solution in the limit of $\Delta\to 0$. 
Using the form of the scaling solution derived for $\Delta\to 0$ in \cite{franz2017} and $z-z_J \sim p$ this matching argument leads to
the scaling behavior of $g(h)$ and $P(f)$ at $\alpha_J$:
\begin{align}
 g(h) \sim \begin{cases}
	    \Delta^{-\mu\gamma}p_0(h\Delta^{-\mu}) & (h\sim \Delta^{\mu}) \\
	    h^{-\gamma} & (h\sim 1)
	   \end{cases},\label{135127_9Jun18}\\
 P(f) \sim \begin{cases}
	    \Delta^{\theta\nu}p_{0}(f\Delta^{-\nu}) & (f\sim \Delta^\nu)\\
	    f^\theta & (f\sim 1)
	   \end{cases},
\end{align}
with new critical exponents $\mu=\kappa/(4\kappa-4)=0.851$,
$\nu=\mu-1/2$, and a universal scaling function $p_0(x)$.  The scaling
analysis also leads to $z_J-1 \sim \Delta^{1/2}$ and $c_\Delta\sim
\Delta^{-1/2}$, consistently with the marginal stability argument,
Eqs.~(\ref{160503_17May18}),~(\ref{155702_18May18}).

The simplicity of the model allows us to derive the analytical form of
the density of states $D(\omega)$. As before, we define the Hessian
matrix as the second derivatives of the interaction potential $V_N$,
Eq.~(\ref{133037_18May18}) w.r.t $x_i$ and $R_\mu/\Delta$. Using the
Edwards-Jones formula for the eigenvalue density
$\r(\l)$~\cite{edwards1976,livan2017}, the density of states
$D(\omega)=2\omega\rho(\omega^2)$ can be expressed analytically in
closed form as a function of $z$, $k_R$ and $p$.  These quantities
should be obtained by solving numerically the full RSB equations but for
simplicity, because here we are interested only in the scaling
properties of $D(\omega)$, to obtain Fig.~\ref{fig:DOS} we used
arbitrary functions $z$, $k_R$ and $p$ which are compatible with the
analytical scaling derived from the full RSB equation.  We find that
$D(\omega)$ displays three separate bands (Fig.~\ref{fig:DOS}).  As in
the standard perceptron~\cite{franz2015}, marginal stability in the full
RSB phase implies that the lowest band starts from $\omega=0$ and for
small $\omega$, $D(\omega)\sim \omega^2$.  The lowest band terminates at
$\omega_0 \sim \D^{1/2}p^{1/2}$ near which $D(\omega)$ exhibits a sharp
peak. At $\omega_1\sim \Delta$ a delta peak is found, while the highest
band starts from $\omega_2 \sim \Delta^{1/2}$.  The qualitative behavior
of $D(\omega)$, and the scaling of $\omega_0$, $\omega_1$ and $\omega_2$
are the same of all the models displaying hypostatic jamming, such as
ellipsoids~\cite{zeravcic2009,schreck2012} and BP~\cite{brito2018}. This
confirms that the generalised perceptron can reproduce analytically all
the critical properties of the hypostatic jamming transition.

\begin{figure}[t]
\begin{center}
 \includegraphics[width=9cm]{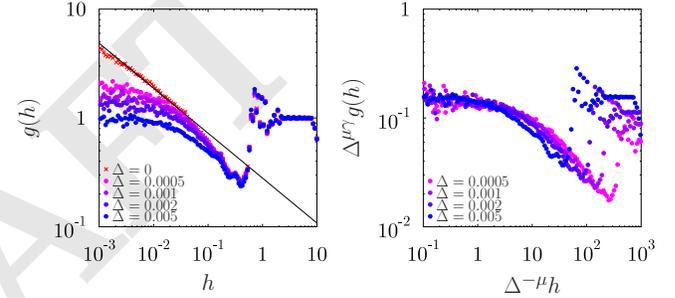} \caption{Gap distribution $g(h)$
 of breathing particles near the jamming point, $p=10^{-6}$. (Left)~Symbols denote
 the numerical result, while the full line denotes the theoretical
 prediction, $g(h)\propto h^{-0.413}$.  (Right)~Scaling plot of the same
 data according to \eqref{135127_9Jun18}.} \label{192110_25May18}
\end{center} 
\end{figure}

As a final check of universality, we test the prediction for the
$\Delta$ dependence of the gap distribution function $g(h)$ at the
jamming point, \eqref{135127_9Jun18}.  In Fig.~\ref{192110_25May18}, we
show numerical results (obtained as in~\cite{brito2018}) for $g(h)$ of
the BP model at $p=10^{-5}$, a value small enough to observe the
critical behavior.  Here, as usual for particle systems, $g(h)$ is
normalized by $g(h)\to 1$ for larger $h$. When $\Delta=0$, $g(h)$
exhibits a power law divergence, $g(h)\sim h^{-\gamma}$, where
$\gamma=0.413$, consistently with previous numerical
observation~\cite{PhysRevE.68.011306,PhysRevE.71.011105,charbonneau2014fractal}.
For finite $\Delta$, on the contrary, the divergence of $g(h)$ is
cutoff (Fig.~\ref{192110_25May18}), consistently with the theoretical prediction of
\eqref{135127_9Jun18}.

\paragraph*{Conclusions --} Using a marginal stability argument, we derived the
scaling behavior of the contact number $z$ and the density of states
$D(\omega)$ of ellipsoids and breathing particles. Our theory predicts
that the scaling behaviors of the two models are identical, which we
confirmed numerically. Many other models of non-spherical particles
display the same jamming criticality~\cite{werf2018}, which defines a
new universality class of hypostatic jamming.  We introduced an
analytically solvable model which allows us
to derive analytically the critical exponents associated to the new
universality class.

One of the most surprising output of our theory is the universality of
the density of states $D(\omega)$ (Fig.~\ref{fig:DOS}). This might be
relevant for some colloidal experiments where the constituents are
non-spherical \cite{kim2006}, in which the vibrational modes could be
experimentally extracted from the fluctuations of
positions~\cite{chen2010, ghosh2010}. Another relevant question is how
non-spherical particles would flow under shear~\cite{mailman2009}. The
divergence of the viscosity at jamming is related to the low
eigenvalues of $D(\omega)$~\cite{lerner2012unified}, which suggests that
the shear flow of non-spherical particles should be quite different from
that of spherical particles, in agreement with recent
experiments~\cite{Tapia2017rheology}.

\acknow{ 
We thank B.~Chakraborty, A.~Ikeda, J.~Kurchan, S.~Nagel and S.~Franz for
interesting discussions. This project has received funding from the
European Research Council (ERC) under the European Union's Horizon 2020
research and innovation programme (grant agreement
n.723955-GlassUniversality).  This work was supported by a grant from
the Simons Foundation (\#454953, Matthieu Wyart and \#454955, Francesco
Zamponi) and by ``Investissements d'Avenir'' LabEx PALM 
(ANR-10-LABX-0039-PALM, P. Urbani).  
We thank the authors of Refs.~\cite{werf2018} and
\cite{schreck2012} for sharing their data used in Figs.~\ref{fig:dz} and
\ref{fig:DOS}.  
}

\showacknow % Display the acknowledgments section

% \pnasbreak splits and balances the columns before the references.
% If you see unexpected formatting errors, try commenting out this line
% as it can run into problems with floats and footnotes on the final page.

%\pnasbreak

% Bibliography
%\bibliography{pnas-sample}
\bibliography{apssamp}

\begin{thebibliography}{10}

\bibitem{Liu10}
Liu AJ, Nagel SR, van Saarloos W, Wyart M (2010) {\em The jamming scenario: an
  introduction and outlook}, eds.{} L.Berthier, Biroli G, Bouchaud J, Cipeletti
  L, van Saarloos W.
\newblock (Oxford University Press, Oxford).

\bibitem{van2009}
Van~Hecke M (2009) Jamming of soft particles: geometry, mechanics, scaling and
  isostaticity.
\newblock {\em Journal of Physics: Condensed Matter} 22(3):033101.

\bibitem{torquato2010}
Torquato S, Stillinger FH (2010) Jammed hard-particle packings: From kepler to
  bernal and beyond.
\newblock {\em Rev. Mod. Phys.} 82(3):2633--2672.

\bibitem{bernal1960}
Bernal J, Mason J (1960) Packing of spheres: co-ordination of randomly packed
  spheres.
\newblock {\em Nature} 188(4754):910.

\bibitem{durian1995}
Durian DJ (1995) Foam mechanics at the bubble scale.
\newblock {\em Phys. Rev. Lett.} 75(26):4780--4783.

\bibitem{PhysRevE.68.011306}
O'Hern CS, Silbert LE, Liu AJ, Nagel SR (2003) Jamming at zero temperature and
  zero applied stress: The epitome of disorder.
\newblock {\em Phys. Rev. E} 68(1):011306.

\bibitem{zhang2009}
Zhang Z, et~al. (2009) Thermal vestige of the zero-temperature jamming
  transition.
\newblock {\em Nature} 459(7244):230.

\bibitem{karayiannis2009}
Karayiannis NC, Foteinopoulou K, Laso M (2009) The structure of random packings
  of freely jointed chains of tangent hard spheres.
\newblock {\em The Journal of chemical physics} 130(16):164908.

\bibitem{donev2004}
Donev A, et~al. (2004) Improving the density of jammed disordered packings
  using ellipsoids.
\newblock {\em Science} 303(5660):990--993.

\bibitem{jaoshvili2010}
Jaoshvili A, Esakia A, Porrati M, Chaikin PM (2010) Experiments on the random
  packing of tetrahedral dice.
\newblock {\em Phys. Rev. Lett.} 104(18):185501.

\bibitem{bi2015}
Bi D, Lopez J, Schwarz J, Manning ML (2015) A density-independent rigidity
  transition in biological tissues.
\newblock {\em Nature Physics} 11(12):1074.

\bibitem{ohern2002}
O'Hern CS, Langer SA, Liu AJ, Nagel SR (2002) Random packings of frictionless
  particles.
\newblock {\em Physical Review Letters} 88(7):075507.

\bibitem{PhysRevLett.97.258001}
Ellenbroek WG, Somfai E, van Hecke M, van Saarloos W (2006) Critical scaling in
  linear response of frictionless granular packings near jamming.
\newblock {\em Phys. Rev. Lett.} 97(25):258001.

\bibitem{PhysRevE.71.011105}
Donev A, Torquato S, Stillinger FH (2005) Pair correlation function
  characteristics of nearly jammed disordered and ordered hard-sphere packings.
\newblock {\em Phys. Rev. E} 71(1):011105.

\bibitem{charbonneau2014fractal}
Charbonneau P, Kurchan J, Parisi G, Urbani P, Zamponi F (2014) Fractal free
  energy landscapes in structural glasses.
\newblock {\em Nat. Commun.} 5:3725.

\bibitem{PhysRevLett.114.125504}
Charbonneau P, Corwin EI, Parisi G, Zamponi F (2015) Jamming criticality
  revealed by removing localized buckling excitations.
\newblock {\em Phys. Rev. Lett.} 114(12):125504.

\bibitem{PhysRevE.72.051306}
Wyart M, Silbert LE, Nagel SR, Witten TA (2005) Effects of compression on the
  vibrational modes of marginally jammed solids.
\newblock {\em Phys. Rev. E} 72(5):051306.

\bibitem{wyart2005rigidity}
Wyart M (2005) On the rigidity of amorphous solids.
\newblock {\em arXiv preprint cond-mat/0512155}.

\bibitem{maxwell1864}
Maxwell JC (1864) L. on the calculation of the equilibrium and stiffness of
  frames.
\newblock {\em The London, Edinburgh, and Dublin Philosophical Magazine and
  Journal of Science} 27(182):294--299.

\bibitem{brito2006}
Brito C, Wyart M (2006) On the rigidity of a hard-sphere glass near random
  close packing.
\newblock {\em EPL} 76(1):149.

\bibitem{PhysRevLett.109.125502}
Wyart M (2012) Marginal stability constrains force and pair distributions at
  random close packing.
\newblock {\em Phys. Rev. Lett.} 109(12):125502.

\bibitem{lerner2013low}
Lerner E, D{\"u}ring G, Wyart M (2013) Low-energy non-linear excitations in
  sphere packings.
\newblock {\em Soft Matter} 9(34):8252--8263.

\bibitem{degiuli2014}
DeGiuli E, Lerner E, Brito C, Wyart M (2014) Force distribution affects
  vibrational properties in hard-sphere glasses.
\newblock {\em Proceedings of the National Academy of Sciences}
  111(48):17054--17059.

\bibitem{franz2015}
Franz S, Parisi G, Urbani P, Zamponi F (2015) Universal spectrum of normal
  modes in low-temperature glasses.
\newblock {\em PNAS} 112(47):14539--14544.

\bibitem{franz2016}
Franz S, Parisi G (2016) The simplest model of jamming.
\newblock {\em Journal of Physics A: Mathematical and Theoretical}
  49(14):145001.

\bibitem{franz2017}
Franz S, Parisi G, Sevelev M, Urbani P, Zamponi F (2017) {Universality of the
  SAT-UNSAT (jamming) threshold in non-convex continuous constraint
  satisfaction problems}.
\newblock {\em SciPost Phys.} 2(3):019.

\bibitem{man2005}
Man W, et~al. (2005) Experiments on random packings of ellipsoids.
\newblock {\em Phys. Rev. Lett.} 94(19):198001.

\bibitem{delaney2005}
Delaney G, Weaire D, Hutzler* S, Murphy S (2005) Random packing of elliptical
  disks.
\newblock {\em Philosophical Magazine Letters} 85(2):89--96.

\bibitem{donev2007}
Donev A, Connelly R, Stillinger FH, Torquato S (2007) Underconstrained jammed
  packings of nonspherical hard particles: Ellipses and ellipsoids.
\newblock {\em Phys. Rev. E} 75(5):051304.

\bibitem{mailman2009}
Mailman M, Schreck CF, O'Hern CS, Chakraborty B (2009) Jamming in systems
  composed of frictionless ellipse-shaped particles.
\newblock {\em Phys. Rev. Lett.} 102(25):255501.

\bibitem{zeravcic2009}
Zeravcic Z, Xu N, Liu A, Nagel S, van Saarloos W (2009) Excitations of
  ellipsoid packings near jamming.
\newblock {\em EPL (Europhysics Letters)} 87(2):26001.

\bibitem{schreck2012}
Schreck CF, Mailman M, Chakraborty B, O'Hern CS (2012) Constraints and
  vibrations in static packings of ellipsoidal particles.
\newblock {\em Phys. Rev. E} 85(6):061305.

\bibitem{williams2003}
Williams SR, Philipse AP (2003) Random packings of spheres and spherocylinders
  simulated by mechanical contraction.
\newblock {\em Phys. Rev. E} 67(5):051301.

\bibitem{blouwolff2006}
Blouwolff J, Fraden S (2006) The coordination number of granular cylinders.
\newblock {\em EPL (Europhysics Letters)} 76(6):1095.

\bibitem{wouterse2007}
Wouterse A, Williams SR, Philipse AP (2007) Effect of particle shape on the
  density and microstructure of random packings.
\newblock {\em Journal of Physics: Condensed Matter} 19(40):406215.

\bibitem{wouterse2009}
Wouterse A, Luding S, Philipse A (2009) On contact numbers in random rod
  packings.
\newblock {\em Granular Matter} 11(3):169--177.

\bibitem{teitel2018}
Marschall T, Teitel S (2018) Compression-driven jamming of athermal
  frictionless spherocylinders in two dimensions.
\newblock {\em Phys. Rev. E} 97(1):012905.

\bibitem{jiao2010}
Jiao Y, Stillinger FH, Torquato S (2010) Distinctive features arising in
  maximally random jammed packings of superballs.
\newblock {\em Phys. Rev. E} 81(4):041304.

\bibitem{delaney2010}
Delaney GW, Cleary PW (2010) The packing properties of superellipsoids.
\newblock {\em EPL} 89(3):34002.

\bibitem{werf2018}
VanderWerf K, Jin W, Shattuck MD, O'Hern CS (2018) Hypostatic jammed packings
  of frictionless nonspherical particles.
\newblock {\em Phys. Rev. E} 97(1):012909.

\bibitem{boromand2018}
Boromand A, Signoriello A, Ye F, O'Hern CS, Shattuk M (2018) Jamming of
  deformable polygons.
\newblock {\em arXiv preprint arXiv:1801.06150}.

\bibitem{baule2013}
Baule A, Mari R, Bo L, Portal L, Makse HA (2013) Mean-field theory of random
  close packings of axisymmetric particles.
\newblock {\em Nat. Commun.} 4:2194.

\bibitem{brito2018}
Brito C, Lerner E, Wyart M (2018) Theory for swap acceleration near the glass
  and jamming transitions.
\newblock {\em arXiv preprint arXiv:1801.03796}.

\bibitem{PhysRevX.7.021039}
Ninarello A, Berthier L, Coslovich D (2017) Models and algorithms for the next
  generation of glass transition studies.
\newblock {\em Phys. Rev. X} 7(2):021039.

\bibitem{alexander1998}
Alexander S (1998) Amorphous solids: their structure, lattice dynamics and
  elasticity.
\newblock {\em Physics reports} 296(2-4):65--236.

\bibitem{gay1981}
Gay J, Berne B (1981) Modification of the overlap potential to mimic a linear
  site--site potential.
\newblock {\em The Journal of Chemical Physics} 74(6):3316--3319.

\bibitem{yan2016}
Yan L, DeGiuli E, Wyart M (2016) On variational arguments for vibrational modes
  near jamming.
\newblock {\em EPL (Europhysics Letters)} 114(2):26003.

\bibitem{schreck2010}
Schreck CF, Xu N, O'Hern CS (2010) A comparison of jamming behavior in systems
  composed of dimer-and ellipse-shaped particles.
\newblock {\em Soft Matter} 6(13):2960--2969.

\bibitem{goodrich2012finite}
Goodrich CP, Liu AJ, Nagel SR (2012) Finite-size scaling at the jamming
  transition.
\newblock {\em Physical review letters} 109(9):095704.

\bibitem{mezard1987}
M{\'e}zard M, Parisi G, Virasoro M (1987) {\em Spin glass theory and beyond: An
  Introduction to the Replica Method and Its Applications}.
\newblock (World Scientific Publishing Company) Vol.{}~9.

\bibitem{edwards1976}
Edwards SF, Jones RC (1976) The eigenvalue spectrum of a large symmetric random
  matrix.
\newblock {\em Journal of Physics A: Mathematical and General} 9(10):1595.

\bibitem{livan2017}
Livan G, Novaes M, Vivo P (2017) Introduction to random matrices theory and
  practice.
\newblock {\em arXiv preprint arXiv:1712.07903}.

\bibitem{kim2006}
Kim JW, Larsen RJ, Weitz DA (2006) Synthesis of nonspherical colloidal
  particles with anisotropic properties.
\newblock {\em Journal of the American Chemical Society} 128(44):14374--14377.

\bibitem{chen2010}
Chen K, et~al. (2010) Low-frequency vibrations of soft colloidal glasses.
\newblock {\em Phys. Rev. Lett.} 105(2):025501.

\bibitem{ghosh2010}
Ghosh A, Chikkadi VK, Schall P, Kurchan J, Bonn D (2010) Density of states of
  colloidal glasses.
\newblock {\em Phys. Rev. Lett.} 104(24):248305.

\bibitem{lerner2012unified}
Lerner E, D{\"u}ring G, Wyart M (2012) A unified framework for non-brownian
  suspension flows and soft amorphous solids.
\newblock {\em Proceedings of the National Academy of Sciences}.

\bibitem{Tapia2017rheology}
Tapia F, Shaikh S, Butler JE, Pouliquen O, Guazzelli E (2017) Rheology of
  concentrated suspensions of non-colloidal rigid fibres.
\newblock {\em Journal of Fluid Mechanics} 827.

\end{thebibliography}

\end{document}